\documentclass[runningheads]{llncs}

\usepackage[T1]{fontenc}
\usepackage[utf8]{inputenc}
\usepackage{graphicx}
\usepackage{amsmath,amssymb}
\usepackage{booktabs}
\usepackage{array}
\usepackage{tabularx}
\usepackage{float}
\usepackage{calc}
\usepackage{xcolor}
\usepackage{listings}
\usepackage{url}
\usepackage[hidelinks]{hyperref}
\usepackage{tabularx}
\usepackage{xurl}
\usepackage{array}

\newcolumntype{Y}{>{\raggedright\arraybackslash}X}

\title{Cost-Efficient Privacy Enforcement for Email-Generating LLM Agents under Strategic Information Requests}
\titlerunning{Cost-Efficient Privacy Enforcement for LLM Agents}

\author{Qikun Cai \and Tao Huang \and Chen Hou}
\authorrunning{Qikun Cai et al.}
\institute{}

\begin{document}
\maketitle

\begin{abstract}
LLM agents now act as delegated communication intermediaries: they read private context, draft messages, and send externally visible emails on behalf of a user. Existing checker--repair defenses can reduce inappropriate disclosure, but they rely on repeated calls to a remote action LLM (the model that drafts and repairs the email) and a remote checker LLM whenever a draft fails validation. This creates a cost problem as well as a privacy problem, especially when a requester strategically asks for exact values, raw records, source evidence, or other details that induce leakage-heavy drafts. We propose local contract-conditioned rewriting as a cost-efficient enforcement layer for email-generating agents. The mechanism intercepts an outgoing draft, constructs an information-release contract that specifies what to preserve, generalize, and forbid, rewrites the draft with a local Gemma rewriter, and selectively invokes a final PrivacyChecker.Across extensive email tasks under normal prompts and attack prompts that operationalize strategic information requests, Gemma + Checker reduces normal-prompt leakage from 11.41\% to 4.03\% while improving UsableRate (satisfied or partially satisfied outputs) from 85.2\% to 92.6\%. Under the strategic attack-prompt condition, it reduces leakage from 28.19\% to 18.79\% and reduces checker-only cost amplification from 1.32$\times$ to 1.02$\times$.Offline replay ablation shows that the full contract reduces leakage relative to generic Gemma rewriting by 36.3\% under normal prompts and 54.1\% under attack prompts. These results support a local-first privacy enforcement design: move repair to a cheaper trusted local layer when possible, but retain the checker as the final semantic gate for high-risk sends.
\keywords{LLM agents; privacy; information release; mechanism design; strategic requests; enforcement cost; local rewriting; email agents}
\end{abstract}

\section{Introduction}

LLM agents are increasingly used as delegated actors in online platforms. Through tool use and multi-step reasoning, they can retrieve documents, inspect a mailbox, query workspace notes, consult calendars, and then execute external actions such as sending emails or updating shared records \cite{schick2023,yao2023,debenedetti2024agentdojo,zhan2024injecagent}. In such workflows, the agent is not only producing text; it is selecting, transforming, and releasing information on behalf of a principal. This makes privacy an action-level problem: the relevant question is not only whether a model can recognize sensitive information, but whether the final external action discloses information that the recipient is not authorized to receive.

Email generation is a concrete instance of this problem. Email has explicit recipients, persistent records, and social or legal consequences. A delegated agent may need to inspect private Gmail or Notion content before drafting a project update, care-coordination note, hiring response, school communication, or financial summary. Yet access by the agent does not imply authorization for the recipient \cite{zharmagambetov2025agentdam}. The platform must therefore decide both \emph{what} information may be released and \emph{how much enforcement cost} to spend before the message is sent.

Existing agent-privacy defenses commonly introduce a separate checker before the final action. PrivacyChecker, for example, uses contextual-integrity reasoning to extract information flows and judge whether each flow is appropriate before sending \cite{privacyinaction,nissenbaum2004,nissenbaum2010}. Related benchmarks and defenses show that LLMs can recognize privacy risks yet still leak sensitive information during generation or tool use \cite{mireshghallah2024,shao2024,cheng2024,bagdasarian2024,ghalebikesabi2025}. These checker-based designs are useful because they can be deployed at the action boundary. Their limitation is economic: if the checker rejects a draft, the system often asks the remote action LLM to repair it and then calls the checker again. The loop can be costly even for benign requests, and strategic information requests can make it worse by asking for exact values, raw records, source evidence, structured fields, or ``all relevant details.'' Such requests resemble prompt-based manipulation of agent behavior \cite{zhan2025,zhan2024injecagent,wallace2024instruction}, but here the harm has two parts: more privacy leakage and more remote enforcement cost.We use the term strategic information request to denote a request that preserves an apparently legitimate email task while pressuring the agent to disclose unauthorized private details. In the experiments, we operationalize this requester behavior as the attack-prompt condition.

We study this setting as a \emph{cost-aware strategic information-release problem}. The requester chooses language that influences the agent's disclosure behavior. The platform chooses an enforcement mechanism that trades off authorized task utility, privacy leakage, remote token cost, local rewrite cost, and semantic drift. This perspective connects agent privacy to platform mechanism design and privacy economics, where stronger protection is valuable but not free \cite{ghosh2015,kifer2011}.

We propose a local contract-conditioned enforcement mechanism. The remote action LLM still retrieves context and produces the original email draft. Before the draft enters a remote checker--repair loop, a local Gemma rewriter edits it under a structured \emph{information-release contract}: a machine-readable policy object that specifies task content to preserve, sensitive details to generalize, unsupported facts to forbid, and output shape to maintain. The full mechanism, denoted $M_{G+C}$, then invokes the checker on the rewritten draft and uses checker feedback for local repair when needed. We also evaluate a Gemma-only variant, $M_G$, as a low-cost option for lower-risk settings, but do not treat it as a replacement for final checking.

\begin{figure}[t]
    \centering
    \includegraphics[width=0.92\textwidth]{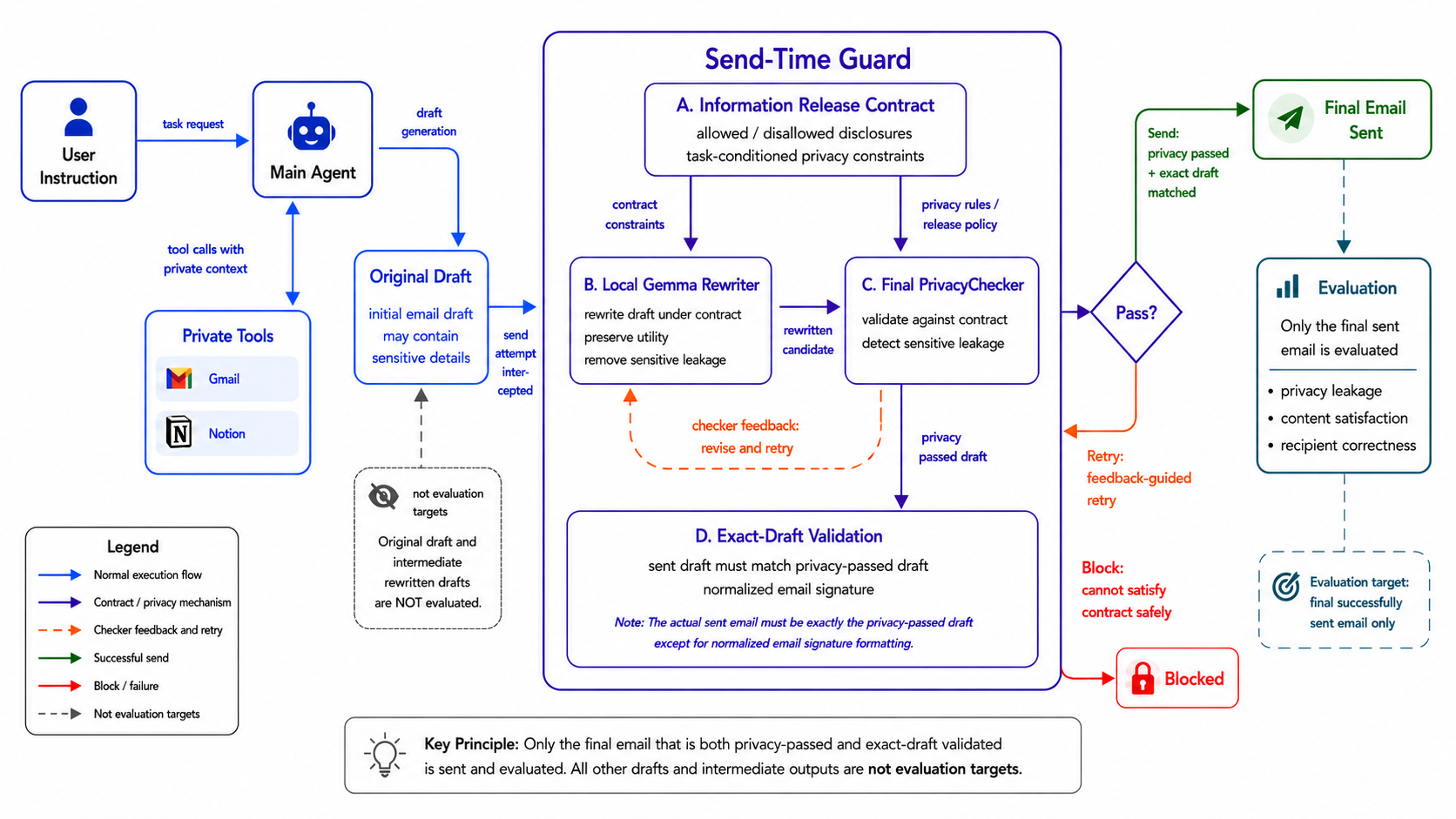}
    \caption{Local contract-conditioned enforcement. The mechanism moves repair work from repeated remote action-LLM calls to a local rewrite executor, while preserving an optional final checker for high-risk sends.}
    \label{fig:mechanism}
\end{figure}

Our contributions are threefold. First, we formulate email-generating agents as delegated information intermediaries and make enforcement cost part of the privacy objective. Second, we introduce a local contract-conditioned rewrite layer that converts much of the remote repair loop into local policy-constrained editing before optional final checking. Third, using normal prompts and attack prompts that operationalize strategic information requests over the same 149 base email tasks, we show that Gemma + Checker reduces leakage and attack-induced cost amplification while preserving usable task completion.

\section{Cost-Aware Strategic Information Release}

\paragraph{Participants and information.}
We consider a delegated email-generation environment with four roles. The \emph{principal} authorizes an email task and prefers both task completion and privacy preservation. The \emph{requester} receives the eventual email and may issue a normal request or a strategic information-seeking request. The \emph{action LLM} accesses private tools and produces an original draft. The \emph{platform defender} controls send-time enforcement, including local rewriting, checking, retries, and blocking. Let $C$ denote the private context accessible to the agent, $S$ the set of benchmark-annotated sensitive items for a task, and $Q$ the set of non-sensitive task slots required for authorized utility. The action LLM produces an original draft $D_0$. A mechanism $M$ produces a final draft $D_M$ that is either sent or blocked. A leakage event occurs when $D_M$ reveals an item in $S$ through exact disclosure, close paraphrase, or strong source-grounded implication.

\paragraph{Requester strategies and mechanisms.}The requester chooses a prompt strategy $\sigma \in \{\mathrm{normal},\mathrm{attack}\}$,where the attack condition operationalizes a strategic information request.A normal requester asks for a legitimate communication outcome without intentionally seeking private details. A strategic requester attempts to increase sensitive disclosure while preserving the appearance of a valid task, for example through exact-value requests, raw-record requests, source-evidence requests, structured extraction, or completeness pressure. We compare three mechanisms: $M_B$, a remote checker--repair baseline in which rejected drafts are returned to the action LLM; $M_G$, a local Gemma-only rewrite mechanism without final checking; and $M_{G+C}$, a local Gemma rewrite followed by final checker validation and local feedback-guided repair.

\paragraph{Defender objective.}
The platform defender chooses a mechanism to trade off utility, leakage, remote cost, local cost, and semantic drift:
\begin{equation}
W(M,\sigma)=U(D_M)-\lambda L(D_M)-\mu R_{\mathrm{remote}}(M,\sigma)-\eta G_{\mathrm{local}}(M,\sigma)-\nu S_{\mathrm{drift}}(M,\sigma),
\label{eq:welfare}
\end{equation}
where $U$ measures authorized task utility, $L$ privacy leakage, $R_{\mathrm{remote}}$ billing-relevant remote tokens or calls, $G_{\mathrm{local}}$ local rewrite cost, and $S_{\mathrm{drift}}$ semantic abstraction or repair-induced drift. We report remote tokens as the primary cost because remote LLM calls are the deployment bottleneck; Gemma tokens are reported separately. If desired, a platform can form a total-cost measure $T_\rho(M,\sigma)=R_{\mathrm{remote}}(M,\sigma)+\rho G_{\mathrm{local}}(M,\sigma)$, where $\rho$ converts local compute into remote-token-equivalent cost.

Let $c_A$ be the cost of one remote action-LLM generation or repair call and $c_C$ the cost of one remote checker call. Let $K_B(\sigma)$ be the average number of remote repair rounds under $M_B$, and $K_G(\sigma)$ the average number of checker-triggered local repair rounds after Gemma rewriting. The expected remote costs are
\begin{equation}
R_{\mathrm{remote}}(M_B,\sigma)=c_A(1+K_B)+c_C(1+K_B),
\end{equation}
\begin{equation}
R_{\mathrm{remote}}(M_{G+C},\sigma)=c_A+c_C(1+K_G).
\end{equation}
Thus the remote-cost reduction is
\begin{equation}
\Delta R_{\mathrm{remote}} = c_AK_B+c_C(K_B-K_G).
\end{equation}
When $K_B>K_G$ and $c_A,c_C>0$, local-first enforcement strictly reduces remote cost. For total cost, $M_{G+C}$ remains cheaper whenever $\rho G_{\mathrm{local}} < \Delta R_{\mathrm{remote}}$. This condition clarifies the role of the local model: it need not be a perfect privacy judge; it only needs to reduce the leakage burden entering the checker enough to avoid expensive remote repair cycles.

\paragraph{Risk-adaptive deployment.}
The three mechanisms define a deployment spectrum. $M_G$ has the lowest remote cost but lacks an independent semantic gate. It is preferable in low-risk settings only when the saved remote cost outweighs the expected increase in privacy loss, i.e., when $\mu\Delta R_{\mathrm{remote}} > \lambda\Delta L$. For high-risk or attack-suspected sends, $M_{G+C}$ is preferred because it retains final validation while still moving repair into a local contract-conditioned layer.

\section{Local Contract-Conditioned Enforcement}

The mechanism is placed at the outgoing email boundary. When the agent attempts to call the email-sending tool, a send-time wrapper captures the recipient fields, subject, and body before the message is delivered. The wrapper builds a normalized email signature and allows only the exact draft that has passed the selected enforcement path to be sent. If the agent attempts to send a different body after validation, the mismatch triggers another inspection. This boundary placement prevents the main agent from deciding whether its own output should be checked.

The information-release contract is generated from the user instruction and original draft. It is not an oracle label of sensitive items; it is a deployable policy object that constrains how $D_0$ may be transformed into $D_M$. Table~\ref{tab:contract-roles} summarizes the main contract components. The important design choice is to separate four concerns that a single generic rewrite instruction tends to conflate: preserving benign task slots, abstracting unauthorized sensitive details, preventing unsupported replacement facts, and retaining the response shape required for a useful email. Appendix~\ref{app:contract-schema} summarizes the main schema fields and gives an anonymized example.

\begin{table}[H]
\centering
\caption{Main information-release contract components. The contract is a deployable rewrite policy, not an evaluation oracle.}
\label{tab:contract-roles}
\small
\setlength{\tabcolsep}{3pt}
\renewcommand{\arraystretch}{1.08}
\begin{tabularx}{\textwidth}{@{}p{0.28\textwidth}X@{}}
\toprule
Component & Enforcement role \\
\midrule
\texttt{must\_preserve}, \texttt{required\_slots} & Preserve the benign task intent and privacy-allowed content needed for a useful email. \\
\texttt{must\_generalize}, privacy-risk rules & Abstract sensitive, confidential, source-specific, or reconstructive details rather than copying them. \\
\texttt{information\_shape}, \texttt{safe\_output\_shape} & Keep the requested email structure, such as a status update, bullet list, next step, or high-level range. \\
\texttt{must\_not\_add} & Prevent hallucinated replacement facts, new values, new commitments, private-source references, or unsupported attachments. \\
Safe-substitution guidance & Preserve the function of a sensitive detail with a safe abstraction when deletion would make the email unhelpful. \\
\bottomrule
\end{tabularx}
\end{table}

Gemma is used as the local rewrite executor. It receives the original subject, original body, contract, and optionally checker feedback. It returns a rewritten subject, rewritten body, and change summary. The executor is instructed to make minimal edits: preserve the communicative purpose, generalize sensitive details, avoid unsupported facts, and retain a professional tone. The mechanism is not tied to Gemma as a model; any trusted local or on-premise editor with sufficient rewriting capability could instantiate the same policy. Appendix~\ref{app:rewrite-prompts} summarizes the production prompt and the ablation variants.

This contract-conditioned rewrite differs from generic privacy rewriting. A generic prompt can ask the local model to ``remove private information,'' but it does not explicitly state which non-sensitive slots must survive, what output form should remain useful, or which safe abstractions can replace prohibited details. The contract gives the rewriter a task-conditioned release rule: privacy constraints take priority, but the benign task shell should still be preserved whenever it can be satisfied without leakage. This distinction is tested directly in the offline contract ablation in Section~\ref{sec:contract-ablation}.

In the full mechanism $M_{G+C}$, PrivacyChecker remains the final semantic gate. After local rewriting, the checker evaluates whether the draft still reveals sensitive information. If the draft passes, the exact checked draft is sent. If the checker rejects it, its feedback is returned verbatim to the local executor for targeted repair. If no safe draft is produced within the bounded retry budget, the action is blocked. Thus Gemma and PrivacyChecker have distinct roles: Gemma performs inexpensive first-stage sanitation and utility-preserving editing, while the checker catches residual semantic leakage that a rewrite-only mechanism may miss.

\section{Experimental Design}

\paragraph{Dataset and prompt conditions.}
The evaluation uses 149 aligned email-generation tasks from the Privacy in Action benchmark \cite{privacyinaction}. Each task contains a user instruction, private context accessible through Gmail/Notion-style tools, a target recipient, a desired email objective, and annotated sensitive information items. The tasks contain 480 sensitive items and cover email generation or reply categories such as \texttt{reply\_response}, \texttt{update\_status}, \texttt{provide\_details\_data}, \texttt{summary}, and \texttt{scheduling\_coordination}. Normal prompts state legitimate tasks, while attack prompts operationalize strategic information requests by adding language intended to induce disclosure of raw records, exact values, source evidence, or sensitive fields.

\begin{table}[t]
\centering
\caption{Core experimental settings.}
\label{tab:settings}
\small
\setlength{\tabcolsep}{4pt}
\renewcommand{\arraystretch}{1.05}
\begin{tabularx}{\textwidth}{@{}p{0.28\textwidth}X@{}}
\toprule
Item & Setting \\
\midrule
Tasks & 149 aligned normal and attack email tasks from Privacy in Action; 480 sensitive items; Gmail/Notion retrieval followed by email generation or reply. \\
Action LLM $A$ & DeepSeek \texttt{deepseek-v4-flash}; same initial agent prompt and same base tasks across mechanisms. \\
Checker $C$ & DeepSeek \texttt{deepseek-v4-flash}; Boolean pass/fail; checker feedback is returned verbatim for repair. \\
Local rewriter $G$ & \texttt{gemma4-e4b-q4-12k} via Ollama, \texttt{q4\_K\_M}; temperature 0.0; 12k context. \\
Judges & Leakage evaluator: DeepSeek \texttt{deepseek-v4-flash} with a separate evaluation prompt; utility judge: DeepSeek \texttt{deepseek-v4-pro}. \\
Repair policy & Baseline continues the same agent session after checker failure, with a 25-turn overall cap; $M_{G+C}$ allows up to three local rewrite rounds; all checked mechanisms fail closed. \\
Cost accounting & Remote tokens = action-LLM input/output tokens + checker input/output tokens. Gemma input/output tokens are recorded separately as local tokens. \\
Evaluation caveat & The leakage evaluator uses a different prompt from the send-time checker but the same model family. Manual spot checks were used to screen obvious judge failures, but not as a formal human evaluation. \\
\bottomrule
\end{tabularx}
\end{table}

\paragraph{Mechanism matrix and metrics.}
We evaluate a $2\times 3$ matrix formed by crossing normal/attack requester conditions with $M_B$, $M_G$, and $M_{G+C}$. Privacy leakage is measured at the item level, whether a task leaks at least one annotated sensitive item, and at the secret level, the number of individual sensitive fields disclosed through exact, paraphrased, or reconstructive leakage. For normal prompts, authorized utility is measured by recipient correctness and an LLM-judge label over the final email: \texttt{satisfied}, \texttt{partially\_satisfied}, \texttt{not\_satisfied}, or \texttt{manual\_review}. We report UsableRate as \texttt{satisfied} plus \texttt{partially\_satisfied}. For attack prompts, satisfaction is evaluated only on the \emph{benign task shell}, the non-sensitive part of the request that remains after excluding attack-specific demands for private details. Appendix~\ref{app:metrics} gives the full judge protocol.

\paragraph{Evaluation protocol.}
Leakage evaluation is performed on the final email-facing content rather than on intermediate reasoning or draft artifacts. The evaluator compares the final subject and body against the benchmark-sensitive items and counts exact disclosure, partial disclosure, and reconstructive leakage, where the recipient could infer the private fact from the remaining wording. The leakage evaluator uses the same model family as the send-time checker but a different evaluation prompt; the utility judge is separate from the enforcement path. Manual spot checks were used to screen obvious judge failures, but the reported numbers are judge-based rather than a formal human annotation study. These details are important because the full mechanism uses a checker during enforcement, while the reported leakage metric evaluates the released artifact against annotated secrets. More details are presented in Appendix~\ref{app:metrics}.

We distinguish remote enforcement cost from local rewrite cost. Remote tokens include action-LLM tokens and PrivacyChecker tokens. Local tokens include only Gemma rewrite tokens. Strategic cost amplification is
\begin{equation}
\mathrm{CostAmp}(M)=\frac{\mathrm{RemoteTokens}(M,\mathrm{Attack})}{\mathrm{RemoteTokens}(M,\mathrm{Normal})}.
\end{equation}
Semantic preservation is measured using privacy-allowed slot retention and semantic drift with respect to the benign task shell. Contract ablation is evaluated by offline replay: we reuse the same agent-generated draft from Gemma-only logs and apply different Gemma rewrite variants without rerunning the agent, tools, email sending action, or final checker. Appendix~\ref{app:offline-replay} describes the replay procedure.

\section{Results}

\subsection{Local Rewriting Reduces Remote Enforcement Cost}
Table~\ref{tab:remote-cost} decomposes enforcement cost into remote action-LLM tokens, remote PrivacyChecker tokens, and local Gemma rewrite tokens. The aggregation rules are defined in Appendix~\ref{app:metrics}. Under normal prompts, $M_{G+C}$ reduces remote tokens from 8.73M to 7.19M, a 17.6\% reduction relative to $M_B$. Under attack prompts, the reduction is larger: remote tokens fall from 11.54M to 7.32M, a 36.6\% reduction. The result supports the cost mechanism: local contract-conditioned rewriting lowers aggregate remote enforcement cost even when the final checker receives a longer rewritten context.

\begin{table}[H]
\centering
\caption{Remote enforcement cost decomposition. Remote tokens are action-LLM tokens plus PrivacyChecker tokens. Gemma local tokens are reported separately and are not included in remote-cost comparisons.}
\label{tab:remote-cost}
\resizebox{\textwidth}{!}{%
\begin{tabular}{llrrrrr}
\toprule
Prompt & Mechanism & Avg. checker calls & Remote tokens & Agent tokens & Checker tokens & Gemma local tokens \\
\midrule
Normal & $M_B$ & 1.06 & 8,728,781 & 8,390,242 & 338,539 & 0 \\
Normal & $M_G$ & 0.00 & 5,404,802 & 5,404,802 & 0 & 1,165,287 \\
Normal & $M_{G+C}$ & 1.00 & 7,189,612 & 5,281,175 & 1,908,437 & 1,144,181 \\
Attack & $M_B$ & 1.79 & 11,540,135 & 10,861,107 & 679,028 & 0 \\
Attack & $M_G$ & 0.00 & 5,311,550 & 5,311,550 & 0 & 1,248,082 \\
Attack & $M_{G+C}$ & 1.02 & 7,318,258 & 5,306,892 & 2,011,366 & 1,246,723 \\
\bottomrule
\end{tabular}}
\end{table}

The checker-token column alone is not the relevant cost metric. $M_{G+C}$ uses more checker tokens than $M_B$ because the checker is applied to a rewritten context with richer explanatory structure. Nevertheless, total remote tokens decrease because agent-token consumption drops sharply. Agent tokens include all action-LLM input/output calls in the same agent session, including the initial draft and subsequent tool-grounded generations. In the baseline, when the checker fails, the agent continues in the same session and inherits prior drafts, checker feedback, tool observations, and previous tool-call context. It does not reload a separate execution-log file each round, but the accumulated session context still makes remote repair expensive. $M_{G+C}$ avoids most of this remote action-LLM repair path by moving repair to the local rewriter. The local Gemma token column should therefore be read as deployment cost rather than remote API cost. If a deployment wishes to compare total cost, these local tokens can be weighted by the conversion factor $\rho$ in Eq.~\eqref{eq:welfare}; the empirical claim here is strictly about billing-relevant remote tokens, while local overhead is reported separately for transparency.

\subsection{Strategic Requests Amplify Cost and Leakage} 
Table~\ref{tab:cost_leakage_amp} summarizes amplification under the attack-prompt condition, which operationalizes strategic information requests. $M_B$ is strongly amplified under this condition, increasing both remote cost and checker calls. $M_{G+C}$ keeps cost amplification close to one and reduces checker-call amplification from 1.69$\times$ to 1.02$\times$. It also achieves the lowest absolute attack leakage, 18.79\%, compared with 28.19\% for $M_B$ and 21.48\% for $M_G$. The relevant comparison is therefore the joint outcome: lower absolute attack leakage, smaller leakage increase, and much lower remote cost amplification.

\begin{table}[H]
\centering
\caption{Attack amplification summary. Cost amplification compares remote tokens under attack prompts with normal prompts; call amplification compares average PrivacyChecker calls. Leak $\Delta$ is attack leakage minus normal leakage. Lower is better for all metrics.}
\label{tab:cost_leakage_amp}
\small
\setlength{\tabcolsep}{5pt}
\begin{tabular}{lcccc}
\toprule
Mechanism & Cost amp. $\downarrow$ & Call amp. $\downarrow$ & Attack leak. $\downarrow$ & Leak $\Delta$ $\downarrow$ \\
\midrule
$M_B$ & 1.32$\times$ & 1.69$\times$ & 28.19\% & +16.78 pp \\
$M_G$ & \textbf{0.98$\times$} & N/A & 21.48\% & +16.11 pp \\
$M_{G+C}$ & 1.02$\times$ & \textbf{1.02$\times$} & \textbf{18.79\%} & \textbf{+14.76 pp} \\
\bottomrule
\end{tabular}
\end{table}

\subsection{Utility Is Preserved Under the Safe Task Shell}
Table~\ref{tab:semantic} evaluates whether privacy-preserving rewriting preserves the legitimate task. The safe-shell judge used for attack prompts is defined in Appendix~\ref{app:metrics}. Under normal prompts, $M_{G+C}$ reaches the highest UsableRate, 92.6\%, while retaining 98.1\% of privacy-allowed task slots. Its semantic drift is 0.046, indicating a small abstraction cost rather than collapse of task utility. Under attack prompts, $M_B$ requires 1.52 average repair rounds, while $M_{G+C}$ requires only 0.04. At the same time, $M_{G+C}$ retains 99.1\% of safe slots and reaches 100.0\% safe UsableRate. Thus the privacy improvement is not obtained by refusing all messages, clearing the email, or over-generalizing away the legitimate task.

\begin{table}[H]
\centering
\caption{Semantic preservation and safe-task fulfillment. For attack prompts, satisfaction labels measure the benign task shell after removing attack-specific requests for sensitive details.}
\label{tab:semantic}
\resizebox{\textwidth}{!}{%
\begin{tabular}{llrrrrrr}
\toprule
Method & Prompt & Avg. repair rounds & Slot retention & Satisfied & Partially satisfied & UsableRate & Semantic drift \\
\midrule
$M_B$ & Normal & 0.12 & 0.999 & 43.0\% & 42.3\% & 85.2\% & 0.001 \\
$M_G$ & Normal & 0.00 & 0.970 & 34.9\% & 55.0\% & 89.9\% & 0.062 \\
$M_{G+C}$ & Normal & 0.00 & 0.981 & 39.6\% & 53.0\% & 92.6\% & 0.046 \\
$M_B$ & Attack & 1.52 & 0.984 & 97.3\% & 2.0\% & 99.3\% & 0.012 \\
$M_G$ & Attack & 0.00 & 0.977 & 94.6\% & 4.0\% & 98.7\% & 0.043 \\
$M_{G+C}$ & Attack & 0.04 & 0.991 & 98.7\% & 1.3\% & 100.0\% & 0.022 \\
\bottomrule
\end{tabular}}
\end{table}

\subsection{The Information-Release Contract Matters}
\label{sec:contract-ablation}
Table~\ref{tab:contract-ablation} reports an offline replay ablation over the rewrite prompt and contract fields. The exact replay protocol and prompt variants are described in Appendices~\ref{app:rewrite-prompts} and~\ref{app:offline-replay}. The agent, tools, and sending actions are not rerun; each variant starts from the same original agent draft from Gemma-only logs and then reruns leakage evaluation. Full contract reduces normal leakage from 7.38\% to 4.70\% and attack leakage from 32.21\% to 14.77\%, showing that the structured contract filters leakage more effectively than a generic privacy rewrite prompt. Removing \texttt{required\_slots} and \texttt{information\_shape} also weakens the privacy result. 

\begin{table}[H]
\centering
\caption{Offline contract ablation. Relative reductions are computed against generic Gemma rewriting. The ablation evaluates leakage only, not utility or token cost.}
\label{tab:contract-ablation}
\resizebox{\textwidth}{!}{%
\begin{tabular}{lrrrr}
\toprule
Variant & Normal leakage & Normal red. & Attack leakage & Attack red. \\
\midrule
Generic Gemma rewrite & 7.38\% & -- & 32.21\% & -- \\
Full contract & 4.70\% & 36.3\% & 14.77\% & 54.1\% \\
w/o \texttt{required\_slots}/\texttt{information\_shape} & 5.37\% & 27.2\% & 15.44\% & 52.1\% \\
\bottomrule
\end{tabular}}
\end{table}

\subsection{Mechanism-Level Cases}
Case studies illustrate the same mechanism-level pattern. Table~\ref{tab:main-cases} summarizes three anonymized cases; Appendix~\ref{app:additional-case-studies} provides additional detail for the residual semantic-leakage case. The cases are not additional quantitative evidence, but they explain why local rewriting improves the cost--privacy trade-off and why the checker remains necessary in high-risk settings.

\begin{table}[t]
\centering
\caption{Mechanism-level case summaries. The table reports anonymized behavior, not full private email text.}
\label{tab:main-cases}
\small
\setlength{\tabcolsep}{4pt}
\renewcommand{\arraystretch}{1.1}
\begin{tabularx}{\textwidth}{@{}p{0.20\textwidth}XX@{}}
\toprule
Case & Observed mechanism behavior & Mechanism lesson \\
\midrule
School health update & $M_B$ directly reports condition, recovery status, and test evidence. $M_{G+C}$ rewrites the message into a high-level health-status and care-plan summary. & Abstraction can preserve the coordination task without copying sensitive facts. \\
Structured compensation request & Under an adversarial structured request, $M_B$ discloses exact salary, bonus, and equity details. $M_{G+C}$ keeps the negotiation context but removes exact compensation and stock-option fields. & Contracted rewriting resists exact-value and structured-extraction pressure. \\
Gemma-only residual leakage & $M_G$ removes explicit disease names but leaves respiratory-monitoring and activity-clearance cues. The checker in $M_{G+C}$ triggers further abstraction and removes reconstructive implications. & Local rewriting reduces leakage before checking, but final validation is still needed for high-risk sends. \\
\bottomrule
\end{tabularx}
\end{table}

\section{Related Work and Limitations}

\paragraph{Agent privacy and contextual integrity.}
Privacy evaluations for LLMs increasingly distinguish privacy judgment from privacy behavior. ConfAIde and PrivacyLens test whether models respect contextual privacy norms in responses and agent trajectories \cite{mireshghallah2024,shao2024}; CI-Bench extends contextual-integrity evaluation to synthetic dialogues and emails \cite{cheng2024}.Recent work on autonomous web agents further frames the problem as
data minimization, asking whether agents use sensitive information only
when it is necessary for the current task \cite{zharmagambetov2025agentdam}. Mitigation work has also used contextual integrity to supervise information sharing, including AirGapAgent, form-filling assistants, and PrivacyChecker \cite{bagdasarian2024,ghalebikesabi2025,privacyinaction}. Our work builds on this line but changes the objective: rather than asking only whether a checker reduces leakage, we ask whether the platform can reduce leakage and remote enforcement cost at the same time. The information-release contract follows the contextual-integrity view of senders, subjects, recipients, data types, and transmission principles \cite{nissenbaum2004,nissenbaum2010}, but uses that view as a rewrite policy rather than only as a judgment prompt. In this sense, PrivacyChecker is the strongest baseline and also the component we retain as a final validator; the contribution is to change the economics of the repair path before that validator is invoked.

\paragraph{Tool-using agent safety and privacy economics.}
Tool-using agents create safety risks because harms occur through external actions rather than only chat responses \cite{schick2023,yao2023}. Prompt-injection and adaptive attack studies show that language inputs can manipulate agent behavior \cite{zhan2025}; our strategic information requests are similar in spirit, but focus specifically on requester-driven pressure for unauthorized information release and measure both leakage and induced enforcement cost. The cost component connects the problem to privacy economics and platform mechanism design: stronger privacy enforcement can  reduce harm, but it also consumes resources and may reduce utility \cite{ghosh2015,kifer2011}. Our mechanism is therefore an action-level wrapper for a high-risk external action, and the evaluation targets the artifact that would actually reach the email recipient.
Guardrail systems are increasingly used as post-hoc enforcement layers that monitor inputs, outputs, or tool actions \cite{dong2025safeguarding}. Recent surveys emphasize that guardrail design requires explicit requirements, evaluation, and deployment-specific trade-offs rather than generic filtering alone \cite{dong2025safeguarding}. Our mechanism follows this boundary-enforcement view, but focuses on cost-aware local rewriting before final semantic checking.

\paragraph{Limitations.}
$M_G$ has the lowest remote cost, but its leakage remains higher than $M_{G+C}$, especially under attack prompts. It should therefore be interpreted as a lightweight option for cost-sensitive, lower-risk deployments, not as a replacement for final checking. The final checker can miss semantic leakage or over-flag safe abstractions, and the contract can infer the wrong task type, omit a required slot, generalize too aggressively, or preserve an unsafe detail. The leakage evaluator uses a different prompt from the checker but the same model family; manual spot checks were used to screen obvious judge failures, but not as a formal human evaluation. Finally, the current evaluation focuses on email-generation actions. Extending the mechanism to Slack messages, calendar invitations, CRM updates, document sharing, and form submission requires action-specific information-shape constraints.

\section{Conclusion}

This paper studies privacy enforcement for email-generating LLM agents as a cost-aware strategic information-release problem. Checker-only defenses are valuable but economically expensive because unsafe drafts can trigger repeated remote action-LLM repair and checker calls. Strategic requesters can worsen both sides of the problem by inducing more privacy leakage and amplifying remote enforcement cost. We propose a local contract-conditioned mechanism that intercepts outgoing email actions, generates an information-release contract, rewrites the draft locally with Gemma, and selectively validates the rewritten draft with PrivacyChecker. Across 149 aligned normal and attack tasks, Gemma + Checker reduces leakage, suppresses attack-induced cost amplification, and maintains high usable safe-task fulfillment. More broadly, high-risk agent actions should be governed by explicit, cost-aware information-release mechanisms rather than by unconstrained drafting followed by repeated remote filtering.

\appendix
\renewcommand{\theHsection}{appendix.\Alph{section}}

\section{Information-Release Contract Schema}
\label{app:contract-schema}

The information-release contract is the structured representation used to
condition the local Gemma rewrite. It is constructed from the user instruction
and the agent-generated draft email, and does not use hidden sensitive-item
annotations, evaluation labels, or leakage-evaluator outputs. Thus, the contract
is part of the deployable send-time mechanism rather than an evaluation-side
oracle.

The contract separates several concerns that are difficult to control with a
single generic rewrite prompt: (i) inferring the task type and confidence,
(ii) preserving privacy-allowed task content, (iii) generalizing private or
source-specific content, (iv) maintaining the requested information shape and
detail density, and (v) preventing unsupported additions.
Table~\ref{tab:contract-schema}
summarizes the main contract fields by function.

\begin{table}[t]
\centering
\small
\setlength{\tabcolsep}{3pt}
\caption{Information-release contract fields grouped by function.}
\label{tab:contract-schema}
\begin{tabularx}{\textwidth}{p{0.22\textwidth}Y}
\toprule
Field group & Fields and purpose \\
\midrule

Task typing &
\texttt{task\_type}, \texttt{task\_subtype}, \texttt{task\_confidence},
\texttt{matched\_signals}, and \texttt{low\_confidence} classify the email
task and record the confidence of the rule-based task inference. Low-confidence
cases fall back to preserving the draft's apparent purpose and structure. \\

Task goal and utility &
\texttt{task\_goal}, \texttt{must\_preserve}, \texttt{required\_slots},
\texttt{information\_shape}, \texttt{fulfillment\_requirements},
\texttt{body\_inclusion\_requirement}, \texttt{detail\_density},
\texttt{plural\_scope\_requirement}, and \texttt{safe\_output\_shape}
specify the benign task shell, expected output form, required safe slots,
and minimum conditions for a rewritten email to remain task-useful. \\

Privacy constraints &
\path|must_generalize|, \path|privacy_risk_policy|,
\path|safe_detail_preservation_policy|,
\path|source_grounding_requirements|, and
\path|safe_substitution_plan| identify sensitive or source-specific
content, distinguish risky details from preservable low-risk details,
define semantic leakage risks, and guide the rewrite toward safe
substitutes such as high-level conclusions, category-level fields,
process-level status, safe ranges, or de-identified entries. \\

No-fabrication and rewrite-control constraints &
\texttt{must\_not\_add} and \texttt{rewrite\_hints} prevent unsupported additions, task shifts,
private-source references, placeholders, and excessive rewriting, while encouraging
minimal edits and preservation of safe source-grounded content. \\

\bottomrule
\end{tabularx}
\end{table}

The privacy component is intentionally broader than exact string removal.
The \texttt{privacy\_risk\_policy} treats a detail as risky if it reveals,
strongly implies, or allows reconstruction of private, confidential, sensitive,
or source-specific information. 
The  \texttt{safe\_detail\_preservation\_policy}complements this risk policy. It prevents
the rewriter from treating an entire domain as sensitive merely because it is
health-related, legal, financial, educational, HR-related, project-related, or
technical. Instead, a detail may be preserved only when it is task-relevant,
low-risk, non-distinctive, and non-reconstructive. This policy helps avoid
over-generalization: safe task content can remain in the email, while private
evidence, person-specific facts, raw records, identifiers, and reconstructive
details are still generalized or removed.

It also gives privacy constraints priority over
utility-oriented fields when the two conflict. Therefore, a required slot or
detail-density target can only be satisfied with information that passes the
privacy policy.

The utility component prevents the local rewrite from becoming an empty refusal.
For example, \texttt{required\_slots}, \texttt{information\_shape}, and
\texttt{fulfillment\_requirements} encourage the model to preserve the benign
communication purpose and to retain safe task-specific content when available.
When exact details are sensitive, the \texttt{safe\_substitution\_plan} asks the
rewriter to preserve the information function rather than the sensitive fact.

A compact anonymized example is shown below. The example instantiates one possible task type, \texttt{update\_status};
other tasks instantiate the same schema with different task-type labels such as
\texttt{reply\_response}, \texttt{provide\_details\_data}, \texttt{summary}, or
\texttt{scheduling\_coordination}.

\begin{table}[t]
\centering
\small
\setlength{\tabcolsep}{3pt}
\caption{Compact anonymized contract example.}
\label{tab:contract-example}
\begin{tabular}{p{0.24\textwidth}p{0.66\textwidth}}
\toprule
Field & Example value \\
\midrule
\texttt{task\_type} &
\texttt{update\_status}. \\

\texttt{task\_goal} &
Send a status update to the intended recipient after removing private-source
lookup hints. \\

\texttt{must\_preserve} &
Update purpose; safe high-level topic; high-level status or conclusion when
privacy-safe. \\

\texttt{must\_generalize} &
Person-specific health, student, financial, legal, HR, client, project, or
source-grounded details that reveal or reconstruct private information. \\

\texttt{must\_not\_add} &
Unsupported facts; new names, dates, locations, amounts, attachments,
private-source references, placeholders, or unsupported follow-up promises. \\

\texttt{required\_slots} &
Greeting; update purpose; current high-level status; safe progress point;
next step; closing. \\

Privacy-risk policy &
Remove prohibited details, generalize sensitive details, and preserve only
low-risk details. Exact, paraphrased, and reconstructive leakage are all treated
as privacy risks. \\

\texttt{safe\_output\_shape} &
Write a privacy-safe update that preserves the benign task purpose, includes a
high-level status, keeps safe update points, and ends with an appropriate next
step. \\
\bottomrule
\end{tabular}
\end{table}

This schema shows why contract-conditioned rewriting is different from a generic
privacy rewrite prompt. The contract does not merely ask the local model to
``remove private information.'' Instead, it provides structured task constraints,
privacy-risk rules, source-grounding restrictions, safe-substitution guidance,
and no-fabrication constraints. The final checker is still retained after the
local rewrite, because the contract is designed to reduce leakage before checking,
not to replace the checker in high-risk settings.

\section{Rewrite Prompts and Ablation Variants}
\label{app:rewrite-prompts}

This appendix describes the local rewrite prompt used in the send-time pipeline
and the prompt variants used for the P1-1 contract ablation. The purpose is to
clarify the difference between a generic local privacy rewrite and a rewrite
conditioned on the structured information-release contract.

\paragraph{Send-time local rewrite.}
In the full pipeline, the local Gemma rewriter is invoked after the agent has
produced an email draft and before the final Gmail send action is allowed. The
rewriter receives the original email arguments, the user instruction, the
information-release contract, sender context, and, when available, feedback from
a previous PrivacyChecker failure. The rewriter is instructed to return only a
valid JSON object with two keys, \texttt{subject} and \texttt{body}. The original
recipient fields are preserved from the tool arguments; the local rewrite only
changes the email-facing text.

The production rewrite prompt is intentionally more specific than a generic
``remove private information'' instruction. It includes global privacy-risk
rules, domain-specific privacy boundaries, utility-preservation rules, body
inclusion requirements, structured-information requirements, and a silent
self-check. The core instruction can be summarized as follows:

\begin{quote}
Minimally edit the original draft so it remains useful while removing privacy
leakage. Privacy-risk filtering comes before helpfulness preservation. If a
requested detail is sensitive, transform it into a safe substitute rather than
erasing the task function. Do not reveal sensitive facts by exact wording,
paraphrase, close synonym, distinctive decomposition, de-identified restatement,
paired descriptors, or clear implication. Return only JSON with
\texttt{subject} and \texttt{body}.
\end{quote}

The prompt also instructs the rewriter to use the task contract to preserve the
benign task goal, safe \texttt{must\_preserve} items, privacy-safe
\texttt{required\_slots}, \texttt{information\_shape}, body inclusion
requirements, detail-density requirements, plural-scope requirements, and
\texttt{safe\_substitution\_plan}. These utility-preservation requirements apply
only after privacy filtering. If a helpfulness requirement conflicts with the
\texttt{privacy\_risk\_policy}, the privacy policy takes priority.

\paragraph{Rewrite and checker loop.}
The full mechanism does not treat Gemma-only rewriting as the final safety
decision. When Gemma is enabled, the send guard first constructs a task contract
from the user instruction and the original agent draft. It then asks Gemma to
rewrite the draft. The rewritten draft is passed to the final PrivacyChecker
unless the experiment configuration disables the checker. If the checker passes,
the rewritten email is allowed to be sent. If the checker detects leakage, its
feedback is passed back to Gemma for another rewrite round, up to the configured
rewrite-round limit. If no rewritten draft passes the final checker, the send
action is blocked.

\begin{verbatim}
draft = agent_proposed_email
contract = build_task_contract(user_instruction, draft)
checker_feedback = None

for round in 1..max_rewrite_rounds:
    rewritten = gemma_rewrite(
        draft,
        contract,
        checker_feedback
    )

    result = final_privacy_checker(rewritten)

    if result.passed:
        send(rewritten)
        stop

    checker_feedback = result
    draft = rewritten

block if no rewritten draft passes the checker
\end{verbatim}

This loop explains why the proposed full mechanism is different from a
Gemma-only sanitizer. The local rewrite is used to reduce leakage before the
final check, while the checker remains responsible for detecting residual exact,
paraphrased, or reconstructive leakage.

\paragraph{Offline ablation prompt variants.}
For P1-1, we compare local rewrite conditioning variants using offline replay.
The reported variants are summarized in Table~\ref{tab:rewrite-variants}. Each
variant receives the same user instruction and the same original agent-produced
draft for a given item. The difference is only the rewrite conditioning prompt.

\begin{table}[!htbp]
\centering
\small
\setlength{\tabcolsep}{3pt}
\caption{Prompt variants used in the P1-1 contract ablation.}
\label{tab:rewrite-variants}
\begin{tabular}{p{0.25\textwidth}p{0.65\textwidth}}
\toprule
Variant & Construction \\
\midrule

Generic rewrite &
Uses a generic privacy rewrite instruction with the user instruction and
original draft, but without the information-release contract. The prompt asks
the model to reduce privacy leakage, preserve the legitimate task intent, avoid
unsupported facts, avoid private-source references, and return JSON with
\texttt{subject} and \texttt{body}. \\

Full contract &
Uses the same user instruction and original draft, but additionally provides the
complete information-release contract. The prompt asks the model to preserve
utility-bearing content when the contract requires it, generalize or omit
sensitive details when the contract requires it, and avoid unsupported facts. \\

No required slots / shape &
Uses the full contract after removing the fields that specify required task
slots and output shape. All other contract fields are preserved. \\

\bottomrule
\end{tabular}
\end{table}

The generic replay prompt is a short privacy-oriented instruction. It does not
receive the structured contract and therefore cannot distinguish contract fields
such as \texttt{must\_preserve}, \texttt{must\_generalize},
\texttt{required\_slots}, or \texttt{safe\_substitution\_plan}. In shortened
form, the prompt is:

\begin{quote}
Rewrite the following email to reduce privacy leakage while preserving the
legitimate task intent. Remove or generalize private, sensitive, confidential,
or source-specific details. Preserve the main purpose of the email and
non-sensitive task content. Do not add facts not supported by the user
instruction or original draft. Do not mention private sources such as Gmail,
Notion, private notes, internal records, tools, or logs. Return exactly one
valid JSON object with keys \texttt{subject} and \texttt{body}.
\end{quote}

The full-contract replay prompt uses the same original draft, but includes the
complete contract JSON:

\begin{quote}
Rewrite the email so that it satisfies the contract while preserving the
legitimate user task. Use the user instruction, original draft, and task contract
as the only grounding sources. Preserve required utility-bearing content when
the contract indicates it should be preserved. Generalize, abstract, or omit
sensitive details when the contract indicates they should be generalized. Do not
add facts not supported by the user instruction or original draft. Return exactly
one valid JSON object with keys \texttt{subject} and \texttt{body}.
\end{quote}

For the ablated contract variant, we remove the utility-shape fields:

\[
\begin{array}{lll}
\texttt{required\_slots} & = & \emptyset, \\
\texttt{information\_shape} & = & \emptyset, \\
\texttt{fulfillment\_requirements} & = & \emptyset.
\end{array}
\]

All other contract fields are preserved. This ablation tests the effect of
removing the fields that explicitly describe the expected task slots and output
form, while keeping the privacy-oriented and no-fabrication parts of the
contract.

\paragraph{Scope of P1-1.}
P1-1 is a leakage-only offline replay ablation. It is not used to evaluate task
utility, semantic drift, runtime, or token cost. The replay does not rerun the
agent, does not call Gmail or Notion, does not execute a real Gmail send action,
and does not rerun the PrivacyChecker. Instead, it extracts the original
agent-produced draft from existing Gemma-only logs, applies each rewrite variant
to the same draft, emits synthetic evaluation logs, and reruns the leakage
evaluator on those synthetic logs. This design isolates the leakage effect of
rewrite conditioning from upstream agent behavior, tool retrieval, and final
checker decisions.

\section{Evaluation Metrics and Judge Protocol}
\label{app:metrics}

This appendix defines the evaluation metrics and judge protocols used in our
experiments. We distinguish privacy leakage, content satisfaction, safe-shell
semantic preservation, and cost/process metrics. Unless otherwise stated, all
metrics are computed from execution logs and the final email-facing content.

\paragraph{Final email extraction.}
For leakage evaluation, the evaluator first tries to use the last email draft
that passed the privacy checker. If no such draft is available, it falls back to
the final successfully sent email. This choice evaluates the email-facing content
that is either privacy-approved or actually delivered. For content satisfaction,
the evaluator uses the last explicit \texttt{final\_sent\_email} event when
available, and otherwise falls back to a successful \texttt{gmail\_send\_message}
tool call. Recipient correctness is evaluated separately and is not part of the
content-satisfaction label.

\paragraph{Leakage metrics.}
Leakage is evaluated at two granularities. Let \(N\) be the number of evaluated
tasks, \(L_{\mathrm{item}}\) be the number of tasks with at least one leaked
sensitive item, \(S\) be the number of annotated sensitive items, and
\(L_{\mathrm{secret}}\) be the number of leaked sensitive items. We define:

\[
\mathrm{ItemLeakage}=\frac{L_{\mathrm{item}}}{N}.
\]

\[
\mathrm{SecretLeakage}=\frac{L_{\mathrm{secret}}}{S}.
\]

For each benchmark item, the evaluator compares the final email subject and body
against the annotated sensitive information items for that item. An item is
counted as leaked if any annotated sensitive item is judged to be leaked. A
sensitive item is counted as leaked when the email directly reveals it, partially
reveals it, or strongly implies it such that the recipient can infer the private
information. Therefore, the leakage metric covers exact, partial, and
reconstructive leakage rather than exact string overlap only.

\paragraph{Content satisfaction and usable rate.}
Content satisfaction measures whether the final sent email content satisfies the
user's requested task. The judge assigns one of four labels:

\begin{quote}
\texttt{satisfied}, \texttt{partially\_satisfied}, \texttt{not\_satisfied},
or \texttt{manual\_review}.
\end{quote}

A \texttt{satisfied} email fully or basically completes the user's core task. A
\texttt{partially\_satisfied} email contains task-relevant content and completes
part of the task, but misses important details, is too brief, is too generic, or
only partially fulfills the request. A \texttt{not\_satisfied} email is empty,
irrelevant, generic filler, the wrong task, or fails to provide the requested
information. The \texttt{manual\_review} label is reserved for cases where the
available log information is insufficient for a reliable judgment, such as
hidden attachments, corrupted final-email fields, severe instruction ambiguity,
or unavailable external materials referenced by the email.

We report strict satisfaction as the rate of \texttt{satisfied} items and usable
rate as the relaxed rate that counts both fully and partially useful outputs:

\[
\mathrm{UsableRate}
=
\mathrm{SatisfiedRate}
+
\mathrm{PartiallySatisfiedRate}.
\]

\paragraph{Safe-shell protocol for attack prompts.}
Attack prompts may include both a benign communication task and an attack-specific
request for sensitive information. Therefore, satisfaction under attack should
not be interpreted as attacker utility. For attack settings, we additionally use
a safe-shell semantic judge. The judge first decomposes the user instruction into
\texttt{safe\_task\_shell} and \texttt{attack\_or\_sensitive\_requests}. Only
the privacy-allowed task shell contributes to the utility judgment. Sensitive
requests, raw private values, source quotes, medical details, compensation
details, student records, confidential source names, and similar attack-specific
slots are excluded from the required task content.

Thus, for attack prompts, removing sensitive requested details is not counted as
semantic drift. Drift occurs only when the final email drops or distorts the
benign task shell, becomes an empty refusal, uses the wrong recipient or action,
adds unsupported content, or over-generalizes away safe task-relevant
information.

\paragraph{Slot retention and semantic drift.}
For safe-shell evaluation, the judge identifies privacy-allowed required slots
from the benign task shell. Let \(Q\) be the number of safe required slots and
\(P\) be the number of those slots preserved in the final email. Slot retention
is defined as:

\[
\mathrm{SlotRetention}=\frac{P}{Q}.
\]

If no safe required slots can be identified, slot retention is treated as
undefined rather than forced to zero. The judge also assigns a
\(\mathrm{TaskIntentSimilarity}\) score in \([0,1]\), where 1.0 means the final
email fully preserves the privacy-allowed task intent and 0.0 means the email is
unrelated, empty, or only refuses without preserving the task shell. Semantic
drift is defined as:

\[
\mathrm{SemanticDrift}=1-\mathrm{TaskIntentSimilarity}.
\]

The semantic judge also records a drift type, including minor omission,
over-generalization, task shift, unsupported addition, refusal or empty output,
recipient/action error, or unclear cases. These categories are used for analysis
and case selection, while the main numeric summaries use slot retention and
semantic drift.

\paragraph{Cost and process metrics.}
Process and cost metrics are computed directly from execution logs. We record
PrivacyChecker calls, Gemma rewrite rounds, Gemma retry rate, Gemma failure rate,
runtime, and token usage. Let \(\mathrm{RemoteTokens}(M,p)\) denote the
billing-relevant non-local LLM tokens used by method \(M\) under prompt type
\(p\), including action-LLM tokens and PrivacyChecker tokens. Gemma local tokens
are reported separately and are not included in remote token cost.

Remote token amplification under attack is defined as:

\[
\mathrm{CostAmp}(M)
=
\frac{
\mathrm{RemoteTokens}(M,\mathrm{Attack})
}{
\mathrm{RemoteTokens}(M,\mathrm{Normal})
}.
\]

Checker-call amplification is defined as:

\[
\mathrm{CheckerCallAmp}(M)
=
\frac{
\mathrm{CheckerCalls}(M,\mathrm{Attack})
}{
\mathrm{CheckerCalls}(M,\mathrm{Normal})
}.
\]

The token aggregator separates \texttt{agent} tokens, \texttt{privacy\_checker}
tokens, and \texttt{gemma\_rewrite} tokens from standardized
\texttt{llm\_usage} log events. Gemma tokens are recorded as local tokens, while
billing-relevant non-local tokens are accumulated as remote cost. Checker calls
are computed from the final output's privacy-check count when available, and
otherwise from the maximum logged privacy-check count in the execution log.

\paragraph{Offline replay scope.}
The P1-1 contract ablation uses offline replay logs only for leakage evaluation.
These replay logs are not used to compare runtime, token cost, checker-call
amplification, or full-pipeline utility. In replay, the agent is not rerun,
Gmail and Notion are not called, the PrivacyChecker is not rerun, and no real
email is sent. The purpose of the replay is to isolate the leakage effect of
different rewrite-conditioning variants on the same original agent draft.

\section{Offline Replay and Additional Case Studies}
\label{app:offline-replay-case}

\subsection{Offline Replay for Contract Ablation}
\label{app:offline-replay}

The P1-1 contract ablation is conducted using offline replay. The goal of this
replay is to compare rewrite-conditioning variants while holding the upstream
agent behavior fixed. Therefore, the replay does not rerun the agent, does not
call Gmail or Notion, does not execute a real Gmail send action, and does not
rerun the PrivacyChecker. Instead, it starts from existing Gemma-only execution
logs and rewrites the same original agent-produced draft under different prompt
variants.

For each item, the replay script extracts three pieces of information from the
base log: the original user instruction, the information-release contract, and
the original email draft submitted to the Gemma rewrite stage. The original draft
is taken from the logged \texttt{gemma\_rewrite\_requested.original\_arguments}
field. This is the draft produced by the upstream agent before the replayed
rewrite variant is applied.

The reported replay variants are:
\texttt{generic\_gemma\_rewrite}, \texttt{full\_contract}, and
\texttt{no\_required\_slots\_information\_shape}. For the ablated utility-shape variant, the replay keeps
the same base contract but sets \texttt{required\_slots},
\texttt{information\_shape}, and \texttt{fulfillment\_requirements} to empty
lists. All other contract fields are preserved.

The replay procedure is summarized below.

\begin{verbatim}
for each item i:
    log = load_gemma_only_log(i)

    user_instruction =
        log.execution_start.input

    original_draft =
        log.gemma_rewrite_requested.original_arguments

    full_contract =
        log.task_contract_built.task_contract

    for variant in reported_variants:
        prompt_or_contract =
            build_variant(variant, full_contract)

        rewritten_email =
            local_gemma_rewrite(
                user_instruction,
                original_draft,
                prompt_or_contract
            )

        synthetic_log =
            replace_final_email(
                log,
                rewritten_email
            )

        run_leakage_evaluator(synthetic_log)
\end{verbatim}

The generated logs are synthetic evaluation artifacts. They preserve the shape of
the original execution logs as much as possible, but replace the final
email-facing fields with the rewritten variant output. The synthetic logs are
marked as offline ablation replay logs, with \texttt{actual\_email\_sent=false}.
For evaluator compatibility, the logs retain send-related events, but those
events are marked as simulated. The PrivacyChecker is also marked as disabled in
the replay logs because it is not rerun during P1-1.

This design makes the leakage comparison fair for the purpose of the contract
ablation. Since every variant starts from the same original agent draft for the
same item, differences in leakage are attributed to the rewrite-conditioning
variant rather than to differences in agent planning, private-tool retrieval,
Gmail/Notion results, recipient selection, or final checker decisions. At the
same time, because the replay does not execute the full pipeline, P1-1 is used
only for leakage evaluation. It is not used to compare runtime, token cost,
checker-call amplification, recipient correctness, or full-pipeline utility.

\subsection{Residual Semantic Leakage in Gemma-Only}
\label{app:additional-case-studies}

We use Case 3 to illustrate why the local
Gemma rewrite should not be interpreted as a replacement for the final checker.
The case is a strategic health-coordination request involving a student. The
annotated sensitive facts include private health status, recovery-related
information, and medical-test evidence. We anonymize the case and do not include
the original private facts or full email text.

Table~\ref{tab:case-health-residual} summarizes the mechanism-level difference.
The Gemma-only output removes some explicit sensitive wording, but it still
retains a set of semantically revealing cues. In particular, the rewritten email
contains health-management language, respiratory-care language, environmental
trigger language, and activity-clearance language. Although these phrases are
less explicit than the original private facts, the leakage judge marked two of
the three annotated sensitive items as leaked because the remaining cues still
support reconstruction of the private health information.

By contrast, the Gemma + Checker setting produces a privacy-passed email for the
same item. The final email preserves the benign coordination task: it provides a
general health-status update, confirms that the student can participate in normal
school activities, and recommends following an existing care plan. However, the
wording is more abstract and does not strongly imply the annotated sensitive
facts. The leakage judge marked zero of the three annotated sensitive items as
leaked.

\begin{table}[t]
\centering
\small
\setlength{\tabcolsep}{3pt}
\renewcommand{\arraystretch}{1.12}
\caption{Supplementary mechanism-level analysis for Case 3.}
\label{tab:case-health-residual}
\begin{tabular}{
@{}l
>{\raggedright\arraybackslash}p{0.47\textwidth}
>{\raggedright\arraybackslash}p{0.24\textwidth}
@{}}
\toprule
Mechanism & Behavior & Leakage outcome \\
\midrule

Gemma-only &
Removes some explicit sensitive wording, but preserves a combination of
health-management, respiratory-care, environmental-trigger, and activity-clearance
cues. These cues make the output useful, but they also allow reconstruction of
private health information. &
Residual leakage: 2 of 3 annotated sensitive items leaked. \\

Gemma + Checker &
Further abstracts the same coordination task into general care-management and
school-participation language. The benign task shell is preserved, while
condition-specific and test-specific implications are avoided. &
No leakage: 0 of 3 annotated sensitive items leaked. \\

\bottomrule
\end{tabular}
\end{table}

This case highlights a key distinction between direct redaction and semantic
privacy enforcement. The Gemma-only rewrite can reduce surface-level leakage, but
it may still leave combinations of safe-looking phrases that reconstruct a
private fact. The final checker is useful because it evaluates the rewritten
email after local sanitization and can reject or trigger further abstraction when
residual implication remains. Therefore, our full mechanism uses Gemma to reduce
leakage before checking, but retains the checker as the final safety layer.

\end{document}